\documentclass[prb,twocolumn]{revtex4-1}
\usepackage[table, dvipsnames]{xcolor}
\usepackage{graphicx}
\usepackage{amssymb}
\usepackage{dcolumn}
\usepackage{float}
\usepackage{bm}
\usepackage{multirow}
\usepackage{booktabs}
\usepackage[T1]{fontenc}
\usepackage[utf8]{inputenc}
\usepackage{lmodern}
\usepackage{color}
\usepackage{makecell}
\usepackage{tikz}
\usepackage{pgfplots}
\usepackage{mathtools}
\usepackage{physics}
\usepackage{amsmath} 
\usepackage[colorlinks=true,linkcolor=red]{hyperref}

\newcommand\scalemath[2]{\scalebox{#1}{\mbox{\ensuremath{\displaystyle #2}}}}

\newcommand{\mathleft}{\@fleqntrue\@mathmargin0pt}

\newcolumntype{M}[1]{>{\centering\arraybackslash}m{#1}}

\DeclareMathAlphabet{\bi}{OML}{cmm}{b}{it}

\def\be{\begin{equation}}
\def\ee{\end{equation}}
\def\bearr{\begin{eqnarray}}
\def\eearr{\end{eqnarray}}

\begin{document}
	
	\title{Magnetoresistive RAM with \textit{n}-doped AlGaAs/GaAs writing/reading channels}
	\bigskip
	\author{Sushmita Saha, Deepak Sain and Alestin Mawrie}
	\normalsize
	\affiliation{Department of Physics, Indian Institute of Technology Indore, Simrol, Indore-452020, India}
	\date{\today}
	\begin{abstract}
		We show that the tunable gate voltage in \textit{n}-doped AlGaAs/GaAs QW (quantum well) is a key in designing an efficient and ultrafast MRAM (magnetoresistive random access memory). The Rashba spin-orbit coupling in such QWs can be tuned appropriately by the gate voltage to create an intense spin-Hall field which in turns interacts with the ferromagnetic layer of the MRAM through the mechanism of spin orbit torque. The strong spin-Hall field leads to an infinitesimally small switching time of the MRAM. Our proposed MRAM is thus a better alternative to the conventional ferromagnetic/spin-Hall effect bi-layers MRAM for the reason that the switching time can be varied with ease, which is unfeasible in the later. Concisely, not only that this work signals a possibility to design an ultra-fast MRAM, but it also suggests a model to fabricate a tunable switching time MRAM.
\end{abstract}
	
\email{amawrie@iiti.ac.in}
	
\maketitle
	
\begin{figure}[b]
		\includegraphics[width=75.5mm,height=35.5mm]{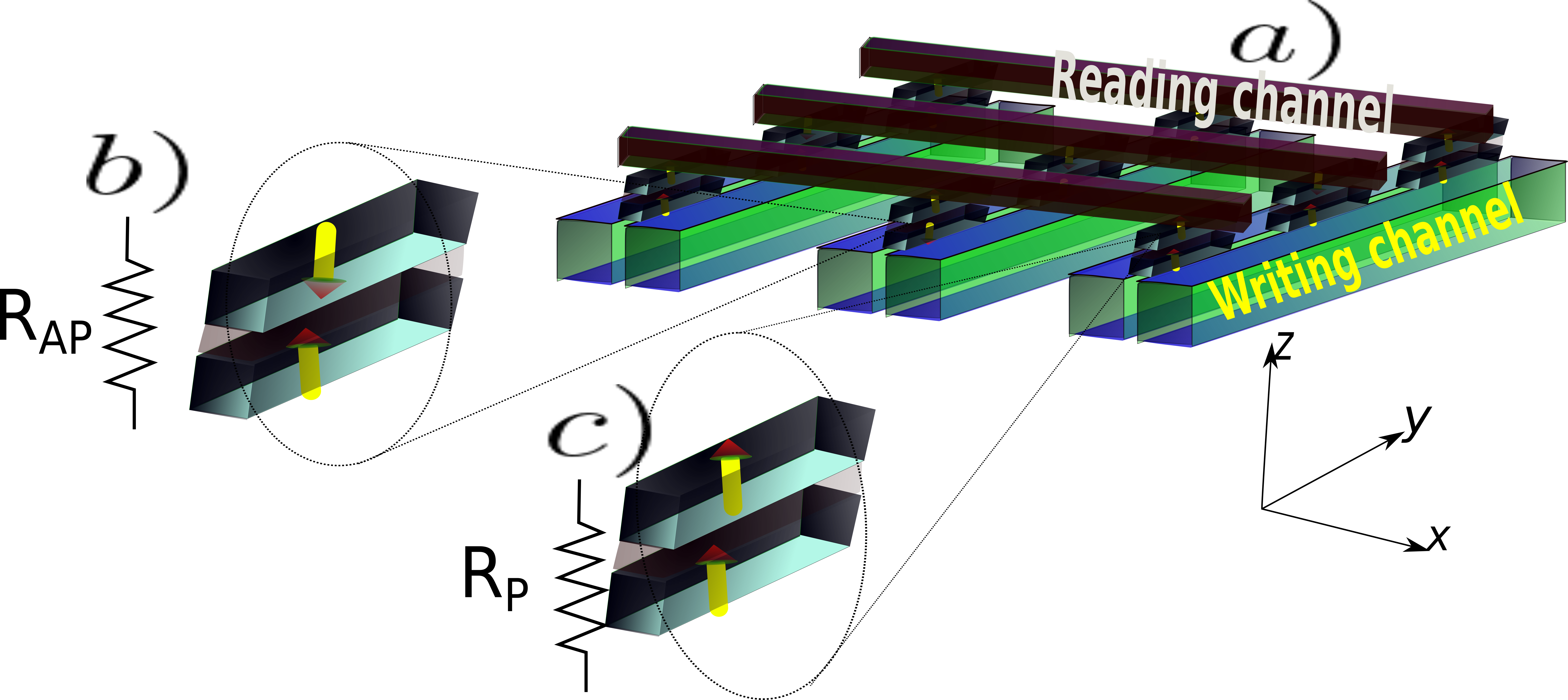}
		\includegraphics[width=50.5mm,height=27.5mm]{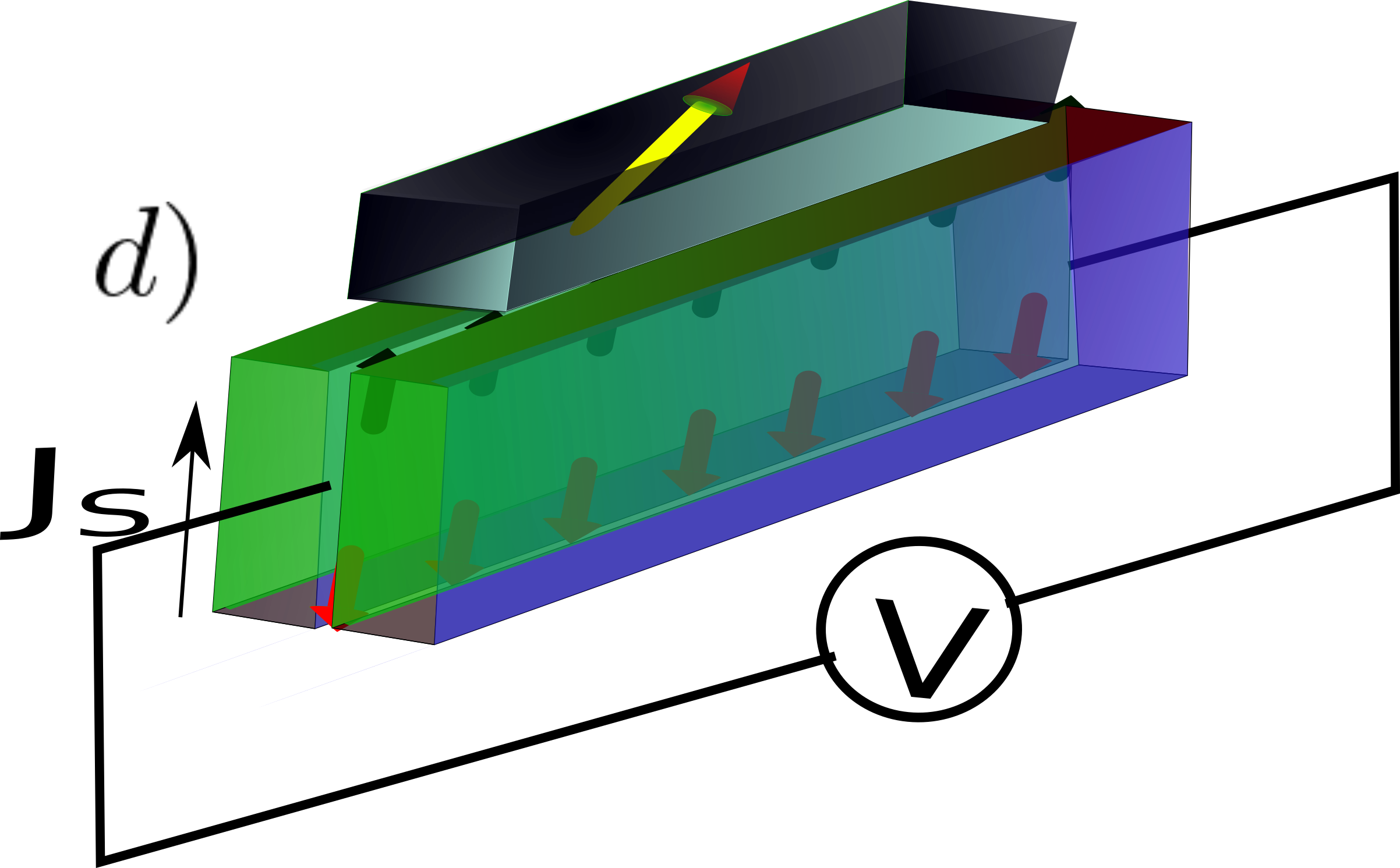}
		\caption{a) Figure showing the schematic of a typical MRAM configuration. Sections b) and c) show the bit at two different states, \textit{viz}, 1 and 0 states, respectively. d) The dominant SHE in the writing node which drives the spin in the free layer (ferromagnetic) of a given bit through the SOT phenomenon.}
		\label{Fig1}
	\end{figure}

\textit{Introduction}: An explicit class of computing memory called magnetoresistive random access memory (MRAM)\cite{mram1,mram2} has recently emerged as a candidate in several integrated device applications. MRAM is non-volatile, low-power-consuming, and fast. What constitutes a single bit of an MRAM are the two ferromagnetic layers separated by a non-magnetic layer (as shown in Fig. [\ref{Fig1} b) \& c)]). The intermediate non-magnetic layer provides a necessary energy barrier to avoid any kind of magnetic interaction between the two ferromagnetic layers. In our design, the magnetization of each layer in a bit can be desirably varied, depending on whether we want to write/read data.  
The traditional way to provide energy for writing/reading data in an MRAM (\textit{viz.} to switch the magnetic state of a ferromagnetic layer) was so far provided by a magnetic field\cite{tr1,tr2}, (which at times becomes too un-scalable) until a recent proposal to do it via the mechanism of spin-orbit torque (SOT)\cite{amin,torejon,cai,fang,mishra,baek,hirai,fan}. 
With a SOT\cite{soc1} based writing/reading channel, an in-plane spin-polarized current from the channel (that interacts with the spin-angular momentum in the ferromagnetic layers) is used to manipulate the magnetization vector in the ferromagnetic layers. The writing channel such as that proposed by V. P. Amin et. al.\cite{amin} are Co/Pt, Co/Cu, and Pt/Cu bilayer set up. One of the important properties of the interface of such bilayers is their ability to exhibit the spin-Hall effect (SHE) phenomenon. The SHE necessitates a flow of an in-plane spin-polarized charge current (see Fig. [\ref{Fig1} d)]) which in turn interacts with a ferromagnetic layer of the bit through the mechanism of SOT.
	
The results presented in this paper originate from the idea of ``replacing the reading/writing channels in Fig. [\ref{Fig1} a)] by the semiconductor heterostructure of \textit{n}-doped AlGaAs/GaAs''. The two-dimensional electron gas (2DEG) trapped between the two compound semiconductors in \textit{n}-AlGaAs/GaAs quantum well is a medium that is well-known for its structural inversion asymmetry (SIA) effect thus providing it, a Rashba spin-orbit coupling (SOC) phenomenon\cite{sia1,sia2,sia3}. 
The SHE in such quantum well is very well pronounced\cite{loss,SHE}. The Rashba SOC in the 2DEG has huge influences on various properties such as electronic transport, magnetotransport, magnetization to name a few\cite{loss1,groshev}, and in this case, it should provide the necessary spin-Hall field to generate a desired magnetization in the ferromagnetic layers of the bit. Additionally, the SOC strength in these 2DEGs can be manipulated by means of a gate voltage\cite{tune1,tune2,tune3,tune4,tune5,tune6,tune7}, which provides a degree of freedom to tune the spin-Hall field. This is unfeasible in the case of a conventional ferromagnetic/SHE bilayers MRAM \cite{amin,torejon,cai,fang,mishra,baek,hirai,fan}, where the amplitude of the spin-Hall field is fixed by the multilayered stack structure.

In this paper, we show that our proposed geometry of the MRAM can function as an ultra-fast MRAM simply by applying a proper tuning of the gate voltage to the writing/reading channels. Specifically, there exists an optimal gate voltage applied to the \textit{n}-doped AlGaAs/GaAs semiconductor heterostructure for which the switching time of the MRAM becomes negligibly small. The reason being, that the spin-Hall field provided by the channel when tuned to such gate-voltage becomes fabulously strong to induce the SOT in the (ferromagnetic/\textit{n}-doped AlGaAs/GaAs) stack structure (as demonstrated in Fig. [\ref{Fig1} c)]).  In the recent experiment\cite{nature}, the 2DEG at oxide interfaces is used as writing/reading channels, however, the discussion on switching time of the MRAM is still lacking. Naturally, the switching time of our proposed MRAM also is gate tunable, which is another piece of a take-home message.

\begin{figure}[http!]
	\includegraphics[width=30.5mm,height=65.5mm]{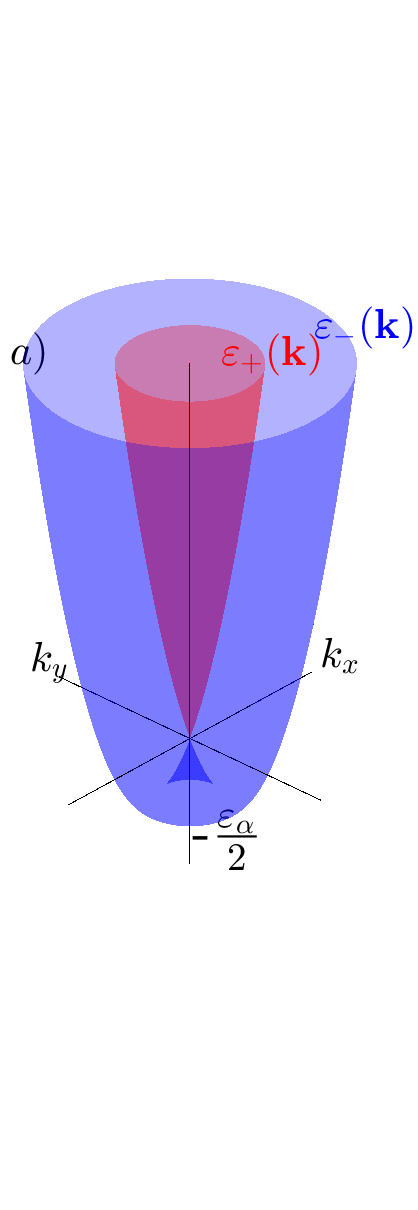}
\includegraphics[width=34.5mm,height=65.5mm]{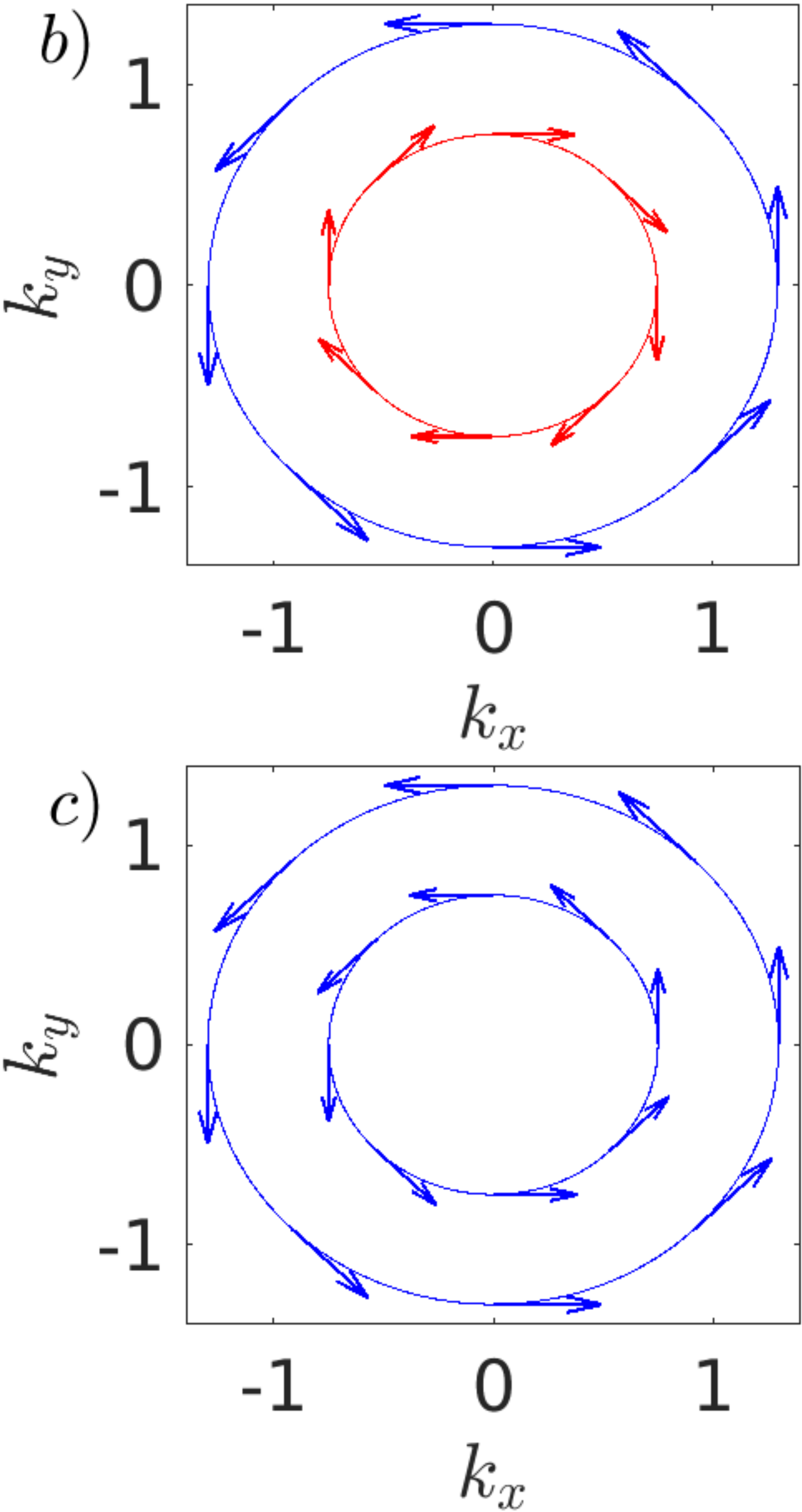}
\caption{a). Plots of the dispersion relation as obtained in Eq. [\ref{disp}]. b) \& c) The fixed energy contour for $\varepsilon_{s}({\bf k})>0$ and $\varepsilon_{s}({\bf k})<0$, indicates the spin-orbit coupling in the material. The two types of arrow (red/blue) denote the two spin states $s=+/-$.}
\label{Fig3}
\end{figure}

\textit{SHE mechanism in the writing/reading channels}:
The single particle Hamiltonian for an electron in the QW of the proposed writing/reading channels is given by\cite{H1, H2,H3} 
\begin{eqnarray}\label{hamil2}
\hat{ H}({\bf p}) & = & \frac{{\bf p}^2}{2m^\ast}+	\frac{\alpha}{\hbar}[\boldsymbol{\sigma}\times{\bf p}]_z,
\end{eqnarray}
Here, $m^\ast$ is the effective mass of the electron in the 2DEG, $\alpha$ is Rashba spin-orbit coupling constant and $\boldsymbol{\sigma}=(\sigma_x,\sigma_y)$ are the $(x,y)$ components of the Pauli's matrices. The eigen-system of the Hamiltonian in the above Eq. [\ref{hamil2}] is given by
\begin{equation}\label{disp}
\varepsilon_s({\bf k})=\frac{\hbar^2k^2}{2m^\ast}+s \alpha k,\;\;
\psi_{n,{\bf k}}({\bf r})=\frac{1}{\sqrt{2}}\begin{pmatrix}
	1\\-s i e^{i\theta}
	\end{pmatrix}e^{i{\bf k}\cdot{\bf r}},
	\end{equation}
with $s=+/-$ denoting the two spin states of the electron. The energy dispersion is demonstrated in Fig. [\ref{Fig3} a)]. The fixed energy contours for energy $\varepsilon_s({\bf k})>0$ and $\varepsilon_s({\bf k})<0$, shown in Fig. [\ref{Fig3} b) \& c)], illustrates the eminent spin-orbit coupling in such QWs.
We also define the lowest energy state  $-\varepsilon_{\alpha}/2$ (refer again to Fig. [\ref{Fig3} a)]), with $\varepsilon_{\alpha}={m^\ast \alpha^2}/{\hbar^2}$. This, in-fact is the gate-voltage tunable parameter\cite{tune1,tune2,tune3,tune4,tune5,tune6,tune7} that we are exploring in this paper. In the subsequent section, we shall see that this energy scale will appear in some experimentally measurable quantities such as the spin-Hall angle and the switching time. It thus becomes a defining energy scale in determining the optimum operational condition of the proposed MRAM (as we shall see in the subsequent sections).
	
\begin{figure}[t]
	\includegraphics[width=33.5mm,height=30.5mm]{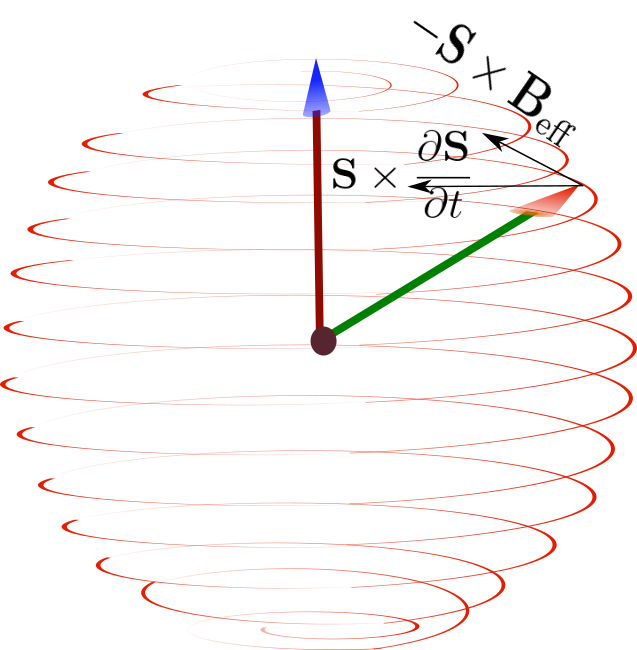}
		\caption{Demonstration of the SOT mechanism between the spin vectors in the writing/reading channel and the corresponding ferromagnetic layer.}
		\label{Fig2}
	\end{figure} 
	
	
Undoubtedly, the main objective here is to study the non-linear evolution of the spin in the ferromagnetic layers, for which we will invoke the \textit{Landau-Lifshitz-Gilbert} (LLG) equation. The spin currents in the writing/reading channel (with spin polarization vector ${\bf s}$) exert a precession-like spin transfer torque of the form $\propto {\bf S}\cross ({\bf S}\cross {\bf s})$ on the spin ${\bf S}$ of the ferromagnetic layer. The ferromagnetic layer also experiences a damping torque of the form ${\bf S}\cross \frac{\partial {\bf S}}{\partial t}$ (see Fig. [\ref{Fig2}]). The LLG equation can be written as a superposition of these two kinds of torque\cite{amin,Ref1,Ref2,Ref3,Ref4,Ref5,Ref6}
\begin{eqnarray}\label{LLG}
\frac{\partial {\bf S}}{\partial t}=-\gamma {\bf S}\cross {\bf B}_{\rm eff}+\beta{\bf S}\cross \frac{\partial {\bf S}}{\partial t}.
\end{eqnarray}
Here, ${\bf B}_{\rm eff}={\bf B}+B_{\rm SH} {\bf S}\cross{\bf s}$ is the effective magnetic field due to the SHE with $\gamma$ being the absolute gyro-magnetic ratio and in the damping term, $\beta$ is the damping constant. The constant $B_{\rm SH}=\frac{\hbar\Theta_{\rm SH} j_e V}{2 e \mu_s t}$, with $\Theta_{\rm SH}$ called as the spin-Hall angle [the ratio of the spin current density due to the SHE and the electrical current density ($j_e$)] and $t$ is the thickness of the ferromagnetic layer.  
For our design of the MRAM, we set the external applied magnetic field ${\bf B}=0$, since we only depend on the spin-Hall field to drive the spin in the ferromagnetic layer of a bit.

The \textit{calculation of the Spin-Hall angle}, ``$\Theta_{\rm SH}$''  is by default based on Eq. [\ref{hamil2} \& \ref{disp}]. 
For the geometry of our problem, we can write $\Theta_{\rm SH}=\sigma_{xy}^z/\sigma_{xx}$, where $\sigma_{xy}^z$ and $\sigma_{xx}$ are respectively, the spin-Hall conductivity and the longitudinal conductivity of the SOC dominant 2DEG in a writing/reading channel. The two mentioned quantities can be written as \cite{JohnC,con1,con2,mahan,spinc}
	\begin{equation}\label{con}
	\left.
	\begin{aligned}
	\scalemath{1}{\sigma_{xy}^z}&=\hbar\iint_{\rm BZ} \frac{\bf dk}{(2\pi)^2}\sum_{s}f_{s,{\bf k}}\\&\times\sum_{s^\prime}\frac{2{\rm Im}[\langle s^\prime,{\bf k}\vert \hat{ J}_x^z\vert s,{\bf k}\rangle \langle s,{\bf k}\vert \hat{ v}_y\vert s^\prime,{\bf k}\rangle]}{(\varepsilon_s({\bf k})-\varepsilon_{s^\prime}({\bf k}))^2}\\
	\scalemath{1}{\sigma_{xx}}&=e^2\sum_{s}\iint_{\rm BZ} \frac{\bf dk}{(2\pi)^2}\Big(-\frac{\partial f}{\partial \varepsilon_{s}({\bf k})}\Big)f_{s,{\bf k}}\\&\times\vert\langle s,{\bf k}\vert \hat{ v}_x\vert s,{\bf k}\rangle\vert^2\tau_{s,{\bf k}},
	\end{aligned}
	\right\} 
	\end{equation}
where $f_{s,{\bf k}}=1/[\exp\{\beta(\varepsilon_{s}({\bf k}))\}+1]$ is the Fermi-Dirac distribution function, the velocity operator $\hat{v}_{x,y}=\partial \hat{H}/\partial p_{x,y}$ in the $x,y$ direction and $\hat{J}_x^z=\{\sigma_z,\hat{v}_x\}$ is the spin-current operator. 
After a rigorous calculation, one can arrive at the following variation of the spin-Hall angle as a function of the chemical potential and the gate tunable energy parameter, $\varepsilon_{\alpha}$
	\begin{eqnarray}\label{sha}
	\Theta_{\rm SH}=
	\begin{cases}
	0. & \text{for } \mu<-\frac{\varepsilon_\alpha}{2}\\
	\frac{\hbar}{2\tau_F\sqrt{\varepsilon_{\alpha}^2+2\mu\varepsilon_{\alpha}}}.& \text{for } \mu < 0\\
	\frac{\hbar}{2\tau_F (\varepsilon_{\alpha}+2\mu)}.              & \text{for } \mu \ge 0
	\end{cases}
	\end{eqnarray}
Here, $\tau_F$ is the relaxation time at the Fermi energy. Note that we have assumed the two spin states have the same relaxation time.
	
	\begin{figure}[t]
		\includegraphics[width=75.5mm,height=50.5mm]{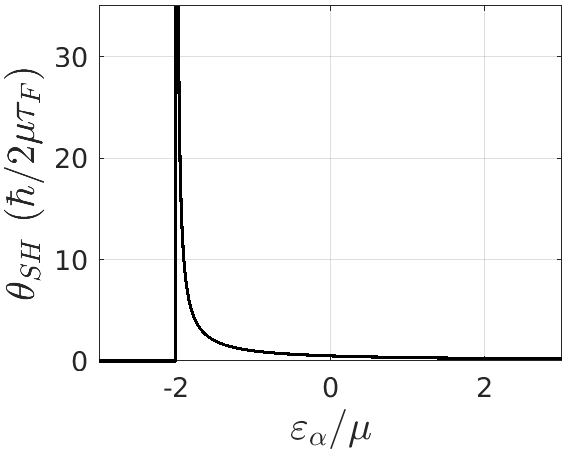}
		\caption{The variation of the spin-Hall angle as a function of the $\varepsilon_{\alpha}/\mu$, illustrating the $\lim_{\varepsilon_{\alpha}\rightarrow -2\mu}\Theta_{\rm SH}=\infty$ condition.}
		\label{Fig4}
	\end{figure} 
From the above Eq. [\ref{sha}], we can see that for a functioning MRAM, the gate voltage applied to the writing/reading channels should be tuned such that $\mu\in[-\frac{\varepsilon_{\alpha}}{2},\infty)$. The variation of the spin-Hall angle as a function of the ratio $\varepsilon_{\alpha}/\mu$, for this kind of range is shown in Fig. [\ref{Fig4}]. We notice that $\lim_{\varepsilon_{\alpha}\rightarrow -2\mu}\Theta_{\rm SH}=\infty$. A large spin-Hall angle indicates a very strong SOT happening between a writing/reading channel and the ferromagnetic layer of the bit. 
We thus set $\varepsilon_{\alpha}\rightarrow -2\mu$ as the optimum operating condition of the MRAM. 
In the next section, we will understand how this desired magnitude of the gate voltage would lead to a very quick switching time, which is desired for an ultra-fast MRAM. Also, it is important to note that the SOT can be turned off completely for an applied gate voltage such that $\varepsilon_\alpha<-2\mu$.

\textit{Calculation of the different components of the magnetization vector from the LLG equation}: Having had  an understanding of the spin-Hall field, we now look into the calculation of the magnetization vector in the ferromagnetic layers by using the LLG equation stated in Eq. [\ref{LLG}]. For our convenience, we rewrite the same LLG equation as
	\begin{eqnarray}
	\frac{\partial {\bf S}}{\partial t}=-\frac{\gamma}{1+\beta^2}{\bf S}\cross {\bf B}_{\rm eff}-\frac{\beta\gamma}{1+\beta^2}{\bf S}\cross({\bf S}\cross {\bf B}_{\rm eff}).
	\end{eqnarray}
Considering the effective magnetic field ${\bf B}_{\rm eff}=B_{\rm SH}{\bf S}\cross{\bf s}$ (as stated earlier) and the spin current in the writing/reading channels polarized about the $z$-axis (as shown in Fig. [\ref{Fig1}]), we yield the following solution for the different components of the spin vector in the ferromagnetic layer:
	\begin{figure}[b]
		\includegraphics[width=62.5mm,height=72.5mm]{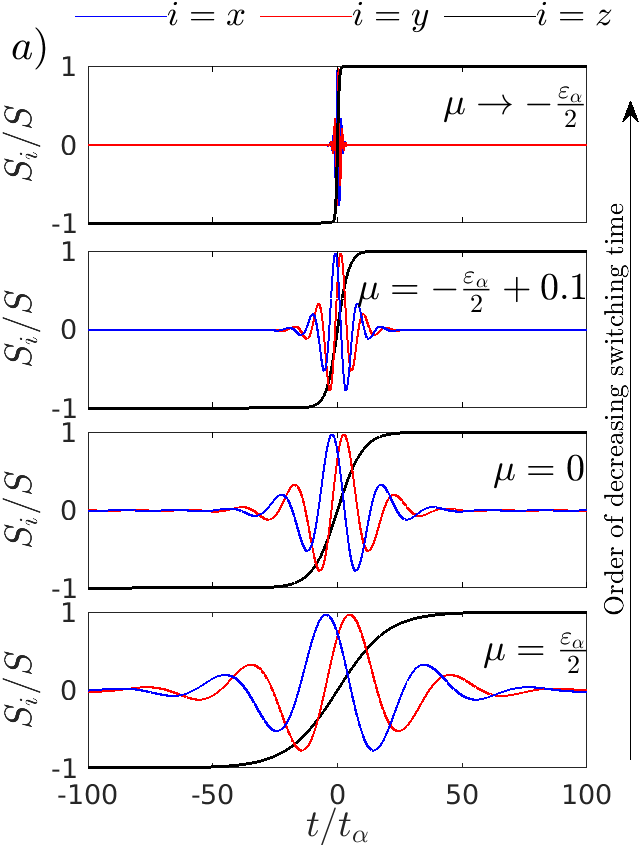}
		\includegraphics[width=42.5mm,height=26.5mm]{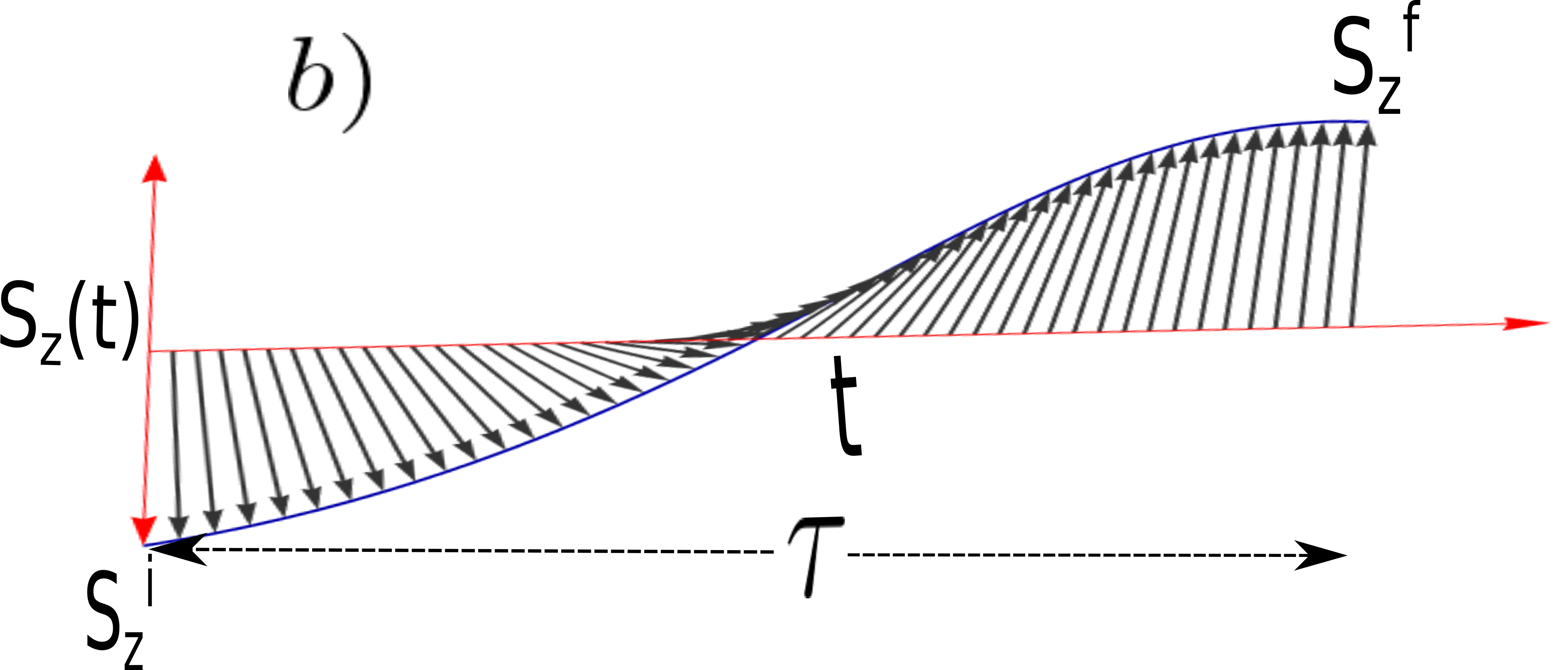}
		\includegraphics[width=40.5mm,height=25.5mm]{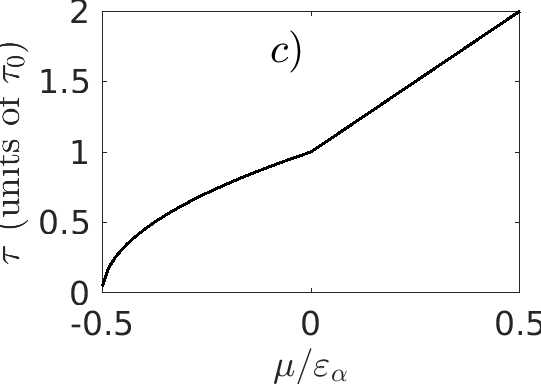}
		\caption{a) The variation $(S_x,S_y,S_z)$ components of the magnetization as a function of time, scaled to $t_\alpha=2\tau_F\theta_0/\hbar\gamma$ for three magnitudes of the gate voltage. b) Evolution of the magnetization vector ${\bf S}=(S_x,S_y,S_z)$ taking the initial value at $t=t_i$, equal to $(0,0,S_z^i)$ and the initial value after $t=\tau$, equal to $(S_x,S_y,S_z)$. c) Plot of the switching time (in units of $\tau_0=\frac{\theta_0\varepsilon_{\alpha}\tau_F}{2\hbar\gamma}(1+\beta^2)\ln\big[\frac{S+S_z^f}{S-S_z^f}\big]$). Also, we define $\theta_0=\frac{2e \mu_s t}{j_e V \hbar}$.}
		\label{Fig5}
	\end{figure} 
	\begin{equation}\label{spin}
	\left.
	\begin{aligned}
	\scalemath{0.95}{\frac{S_x(t)}{S}}=&\scalemath{0.94}{\sech\Big(\frac{S \gamma B_{\rm SH}}{1+\beta^2} t+\phi_1\Big)\sin\Big(\frac{S \gamma \beta B_{\rm SH}}{1+\beta^2} t+\phi_2\Big)} \\
	\scalemath{0.95}{\frac{S_y(t)}{S}}=&\scalemath{0.94}{\sech\Big(\frac{S \gamma B_{\rm SH}}{1+\beta^2} t+\phi_1\Big)\cos\Big(\frac{S \gamma \beta B_{\rm SH}}{1+\beta^2} t+\phi_2\Big)}\\
	\scalemath{0.95}{\frac{S_z(t)}{S}}=&\scalemath{0.94}{\tanh\Big(\frac{S \gamma B_{\rm SH}}{1+\beta^2} t+\phi_1\Big)}
	\end{aligned}
	\right\} 
	\end{equation}
where $S=\sqrt{S_x^2+S_y^2+S_z^2}$ can be regarded as a fixed constant and in our case, we'll take $S=1$. Also, $\phi_1$ and $\phi_2$ are the constant of integration. It is easy to see from our geometry in Fig. [\ref{Fig1}], that the analysis of the component, $S_z(t)$ is sufficient to determine the switching time of the MRAM, which is the all-important measurable quantity. Before that, let us determine the constant of integration, $\phi_1$. We consider the value $S_z(t=0)=S_{z}^0$, which yields 
$\phi_1=\frac{1}{2}\ln \frac{S+S_{z}^0}{S-S_{z}^0}$. The variation of the different components of the spin-vector as a function of time is shown in Fig. [\ref{Fig5} a)], (taking $\phi_2=\pi/4$ and $S_z^0=0$). The different panels in Fig. [\ref{Fig5} a)] are for [$\mu=(\frac{\varepsilon_{\alpha}}{2},-\frac{\varepsilon_{\alpha}}{2}+0.1,-\frac{\varepsilon_{\alpha}}{2})$ from bottom to top]. Clearly, as we approach $\mu\rightarrow-\frac{\varepsilon_{\alpha}}{2}$, the $S_z$ component of the magnetization vector takes little to no time to switch from [$S_z(t=t_i)=-{\rm min\; value}$] to [$S_z(t=t_f)={\rm max\; value}$], which leads us to the concept of switching time of the MRAM. 

We define the switching time of the MRAM as the time required to turn the magnetization of a given bit from the initial state $S_z(t=t_i)=-S_z^i$ (min value) to the desired final state $S_z(t=\tau)=S_z^f$ (max value) (as demonstrated in Fig. [\ref{Fig5} b)]). The time for the spin component $S_z$ to switch from $S_z(t=0)=S_z^0$ to $S_z(t=t_f)=S_z^f$ can be analytically written as\cite{hannay}
	\begin{eqnarray}
	\lambda=\frac{1+\beta^2}{2\gamma B_{\rm SH}}\ln \frac{(S+S_{z}^f)(S-S_{z}^0)}{(S-S_{z}^f)(S+S_{z}^0)}.
	\end{eqnarray}
Clearly, the switching time $\tau=2\lambda$, when we set $S_{z}^0=0$.
The variation of the switching time as a function of the ratio of the chemical potential and energy parameter, $\varepsilon_{\alpha}$ is shown in Fig. [\ref{Fig5} c)]. An ultra-fast MRAM must perceptibly have a negligibly small switching time, strictly speaking, $\tau\rightarrow0$. As evident from the figure, this is easily attainable when the chemical potential is tuned to the limit $\mu\rightarrow -\frac{\varepsilon_{\alpha}}{2}$. Conversely, we can achieve this by taking writing/reading channels with a  fixed doping, and by appropriately varying the gate voltage (also seems more practical from the experiment point of view), thus the parameter, $\varepsilon_{\alpha}$. This luxury of tuning the switching time by means of a gate potential is lacking in the conventional ferromagnetic/spin-Hall effect heterostructure MRAM system, in which the amplitude of the spin-Hall field  at a given current is fixed by the multi-layered stack structure\cite{amin,torejon}. 

	%
	
In \textit{conclusion}, we have proposed an ultra-fast Magnetoresistive RAM by using the SIA \textit{n}-doped AlGaAs/GaAs quantum well as writing/reading channels. The very simple idea is to tune the gate voltage applied to the semiconductor heterostructure appropriately, such as to produce a strong spin-orbit Hall field for initiating the spin-orbit torque between the channel and a ferromagnetic layer of the MRAM. We have theoretically shown that a desired magnitude of the gate-voltage, (such that $\varepsilon_{\alpha}=-2\mu$) brings about a negligibly small switching time of the MRAM. We have also shown that the MRAM has a variable switching time (again, by means of the tunable spin-orbit coupling). We thus propose that the \textit{n}-doped AlGaAs/GaAs semiconductor heterostructure should act as a better alternative to the conventional spin-Hall effect set-up for functioning as a reading/writing channel of an MRAM. In a nutshell, not only that this work signals the possibility to design an ultra-fast MRAM, but it also suggests the possibility of designing a tunable switching time MRAM.

\textit{Acknowledgments}: This work is an outcome of
the Research work carried out under the DST-INSPIRE project DST/INSPIRE/04/2019/000642, Government of India. A. M. also thanks Professor T. K. Ghosh for his valuable inputs.
	
\end{document}